
\input harvmac
\overfullrule=0pt
\newcount\figno
\figno=0
\def\fig#1#2#3{
\par\begingroup\parindent=0pt\leftskip=1cm\rightskip=1cm\parindent=0pt
\baselineskip=11pt
\global\advance\figno by 1
\midinsert
\epsfxsize=#3
\centerline{\epsfbox{#2}}
\vskip 12pt
{\bf Fig. \the\figno:} #1\par
\endinsert\endgroup\par
}
\def\figlabel#1{\xdef#1{\the\figno}}
\def\encadremath#1{\vbox{\hrule\hbox{\vrule\kern8pt\vbox{\kern8pt
\hbox{$\displaystyle #1$}\kern8pt}
\kern8pt\vrule}\hrule}}

\overfullrule=0pt

%

\def\bar{\overline}

\font\zfont = cmss10 
\font\litfont = cmr6

\def\bigone{\hbox{1\kern -.23em {\rm l}}}
\def\ZZ{\hbox{\zfont Z\kern-.4emZ}}
\def\half{{\litfont {1 \over 2}}}

\def\l{\langle}
\def\r{\rangle}

\def\mg1{{\cal M}_{g,1}}
\def\cmg1{{\overline{\cal M}}_{g,1}}

\def\tbar{{\bar t}}
\def\What{{\hat W}}
\def\Wop{{\cal W}^{(n)}}
\def\Wopone{{\cal W}_1^{(n)}}
\def\Woptwo{{\cal W}_2^{(n)}}
\def\Wopg{{\cal W}_g^{(n)}}

\def\Wcal{{\cal W}}
\def\phihat{{\hat \phi}}

\def\np{Nucl. Phys.}
\def\pl{Phys. Lett.}

\def\mpl{Mod. Phys. Lett.}

\def\GINSMOORE{P. Ginsparg and G. Moore, {\it Lectures on 2-D Gravity
and 2-D String Theory}, Los Alamos and Yale Preprint LA-UR-92-3479,
YCTP-P23-92; hep-th/9304011, and references therein.}
\def\LOSEV{A. Losev, Theor. Math. Phys. {\bf 95} (1993) 595.}
\def\DVV{R. Dijkgraaf, H. Verlinde and E. Verlinde, \np\ {\bf B352}
(1991) 59.}
\def\MV{S. Mukhi and C. Vafa, hep-th/9301083, \np\ {\bf B407} (1993) 667.}
\def\GM{D. Ghoshal and S. Mukhi, hep-th/9312189, \np\ {\bf B425}
(1994) 173.}
\def\VAFA{C. Vafa, Mod. Phys. Lett. {\bf A6} (1991) 337.}
\def\DMP{R. Dijkgraaf, G. Moore and R. Plesser, hep-th/9208031,
\np\ {\bf B394} (1993) 356.}
\def\HOP{A. Hanany, Y. Oz and R. Plesser, hep-th/9401030, \np\ {\bf
B425} (1994) 150.}
\def\WITTENGRAV{E. Witten, \np\ {\bf B340} (1990) 281.}
\def\WITTENINT{E. Witten, Surveys in Differential Geometry
{\bf 1} (1990) 243.}
\def\WITTENKS{E. Witten, \np\ {\bf B371} (1992) 191.}
\def\KPZ{V. Knizhnik, A. Polyakov and A.B. Zamolodchikov, \mpl\ {\bf
A3} (1988) 819.}
\def\GROSSKLEB{D. Gross and I. Klebanov, \np\ {\bf B344} (1990) 475.}
\def\EYY{T. Eguchi, Y. Yamada and S.-K. Yang, Preprint UTHEP-275 (May
1994)\hfill\break hep-th/9405106\semi
Y. Lavi, Y. Oz and J. Sonnenschein, hep-th/9406056, \np\ {\bf B431}
(1994) 223.}
\def\OTHER{U. Danielsson, hep-th/9401135, \np\ {\bf B425} (1994) 261\semi
T. Eguchi and H. Kanno, hep-th/9404056, \pl\ {\bf B331} (1994) 330\semi
K. Takasaki, Preprint KUCP-0067 (March 1994), hep-th/9403190\semi
M. Becker, Preprint CERN-TH-7173-94 (March 1994),
hep-th/9403129\semi
K. Becker, Preprint CERN-TH-7174-94 (February 1994),
hep-th/9404157\semi
S. Hirano and K. Ishikawa, hep-th/9406171, \mpl\ {\bf A9} (1994) 3077\semi
L. Bonora and C.S. Xiong, Preprint SISSA-54-94-EP, hep-th/9405004}

{\nopagenumbers
\Title{hep-th/9410034, MRI-PHY/13/94, CERN-TH-7458/94, TIFR/TH/39-94}
{\vbox{\centerline{Topological 2D String Theory:}
\medskip
\centerline{Higher-genus Amplitudes and $W_\infty$ Identities}}}
\ \vskip -1truecm
\centerline{Debashis Ghoshal\foot{E-mail: ghoshal@mri.ernet.in}}
\vskip 4pt
\centerline{\it Mehta Research Institute of Mathematics
\&\ Mathematical Physics}
\centerline{\it 10 Kasturba Gandhi Marg, Allahabad 211 002, India}
\vskip 7pt
\centerline{Camillo Imbimbo\foot{E-mail: imbimbo@vxcern.cern.ch}}
\vskip 4pt
\centerline{{\it Theory Division, CERN, CH-1211 Geneva 23,
Switzerland}\foot{On leave from
{\it INFN, Sezione di Genova, Genoa, Italy}}}
\vskip 7pt
\centerline{Sunil Mukhi\foot{E-mail: mukhi@theory.tifr.res.in}}
\vskip 4pt
\centerline{\it Tata Institute of Fundamental Research,}
\centerline{\it Homi Bhabha Rd, Bombay 400 005, India}
\ \medskip
\noindent We investigate Landau-Ginzburg string theory with the singular
superpotential $X^{-1}$ on arbitrary Riemann surfaces. This theory,
which is a topological version of the $c=1$ string at the self-dual
radius, is solved using results from intersection theory and from the
analysis of matter Landau-Ginzburg systems, and consistency
requirements. Higher-genus amplitudes decompose as a sum of
contributions from the bulk and the boundary of moduli space. These
amplitudes generate the $W_\infty$ algebra.
\vfill
\leftline{September 1994}\eject}
\ftno=0
\newsec{Introduction}
The most interesting solvable string theory corresponds to a two
spacetime dimensional background for the propagation of the bosonic
string. Its solution via matrix models has provided useful information
about string dynamics to all orders in string perturbation
theory\ref\ginsmoore{\GINSMOORE}.

The topological reformulation of this theory has provided an alternate
route to understand the matrix model results, at least for the special
case of compactification of one dimension at the self-dual radius.
The first such reformulation\ref\mv{\MV}, as a topological
Kazama-Suzuki type coset model\ref\wittenks{\WITTENKS}, explained the
origin of the remarkable fact that the genus-$g$ partition function is
proportional to the Bernoulli numbers. In addition, some genus-0
amplitudes could be computed in this approach. This work also
elucidated some facts about the $c=1$ string which had remained
mysterious for some time, such as the existence of the so-called
KPZ\ref\kpz{\KPZ} description based on $SL(2,R)$ current algebra, and
the apparent similarity between the $c=1$ string and coset models
describing two-dimensional black holes.

Subsequently, a different topological reformulation was proposed, in
Ref.\ref\gm{\GM} (and independently in Ref.\ref\hop{\HOP}) in terms of
the infrared fixed-point of a Landau-Ginzburg (LG) type topological
theory, with a superpotential $X^{-1}$. This had the advantage that
genus-0 amplitudes could be computed rather easily and shown to agree
with matrix model results. (For related work, see
Ref.\ref\other{\OTHER}.)  However, the full solution of the theory, in
the very elegant form of $W_\infty$ constraints on the partition
function, was still lacking from the topological approach.

In the present work, we use a combination of results from matter LG
systems in higher genus and intersection theory on the moduli space of
Riemann surfaces, along with consistency requirements, to argue that
this approach gives a topological derivation of $W_\infty$. (Other
approaches to higher-genus have been discussed recently in
Ref.\ref\eyy{\EYY}.)

In particular, this derivation confirms explicitly what had earlier been
argued indirectly: that the topological models really do correspond to
matrix models at the special self-dual value of the radius. Since
genus-0 correlators are insensitive to the radius of compactification,
it is necessary to obtain higher-genus correlators and compare them
with matrix models to directly deduce the value of the radius.

More generally, the importance of a topological derivation of
higher-genus amplitudes is the following. {}From matrix models at the
self-dual radius, one can not only obtain $W_\infty$ identities which
summarize all the tachyon correlators, but also recast them in a
Kontsevich-Penner integral form which is closely related to
topological properties of the moduli space of Riemann
surfaces\ref\dmp{\DMP}. However, it remains quite unclear what is the
fundamental feature of string theory that is responsible for the
emergence of these topological properties, and this is greatly
clarified once the same results are shown to follow from a manifestly
topological reformulation of the string theory itself. In view of the
fact that a topological symmetry algebra underlies all string
backgrounds, this is likely to provide a direction to understand
general topological features of string theory itself, independent of
the background.

\newsec{Review: Amplitudes on the sphere}

Consider the topologically twisted Landau-Ginzburg theory with a
single superfield $X$ and superpotential $W=-1/X$. In the unperturbed
theory, tachyons $T_k$ are given by powers $X^{k-1}$ for $k$ any
integer. Tachyons with $k>0$, called positive tachyons, behave
analogously to the primaries of the conventional polynomial
Landau-Ginzburg models. They can be used to perturb the
superpotential, and generate the so-called small phase space.

In this phase space, the perturbed superpotential $W(X,t)$ and tachyons
$T_k(X,t)$ (where $t$ represents the perturbing parameters $t_k$) are
determined by their flow equations
\eqn\flow{\eqalign{
{\del\over\del t_{k_i}}\; T_{k_j}(X,t) &=
C_{W(X,t)}\left(T_{k_i}(X,t),T_{k_j}(X,t)\right)\cr
{\del\over\del t_{k_i}}\; W(X,t) &=  T_{k_i}(X,t),\qquad\hbox{for }k_i>0,
k_j\in\ZZ\cr}}

Explicit solution of the equations above shows that the tachyons with
positive momenta are constants of the flow, $T_k(X,t)= X^{k-1}$, $k\ge
1$, while the potential is a linear function of the couplings:
\eqn\potential{
W(X,t) =  - X^{-1} + \sum_{k=1}^\infty t_k T_k = - X^{-1} +
\sum_{k=1}^\infty t_k X^{k-1}}
Tachyons with negative momenta $T_{-n}$, $n>0$ can be solved for order
by order in $t$, and in analogy with the polynomial LG theories, can
be expressed as:
\eqn\negatac{
T_{-n}(X,t) = \left({(-W(X,t))^n\over -n}\right)'_-}
where the minus-subscript denotes the terms which are negative powers of
$X$.

The simplest correlator is the genus-0 3-point function $\l\l T_{k_1}
T_{k_2} T_{k_3}\r\r_0$ for which the moduli space is a single point.
The expression for the correlation function is therefore the same as
that in the matter theory. Consider first the case where there
is only one negative momentum tachyon, $k_1=-n, k_2,k_3\ge 0$ and use
the flow equations \flow\ and \negatac\ to write
\eqn\threept{\l\l T_{-n}(t)T_{k_2}T_{k_3}\r\r_0 = \oint {1\over W'}\;
\left({(-W)^n\over -n}\right)'_-\;
{\del W\over\del t_{k_2}}\; {\del W\over\del t_{k_3}} }
We can remove the minus-subscript above since the positive terms do
not contribute to the integral. Then using the fact that the positive
momentum tachyons are constants of the flow, we integrate \threept\
above to write the one-point function of a negative tachyon
$T_{-n}$, $n>0$ (in the small phase space) as
\eqn\onept{ \l\l T_{-n}\r\r_0 =  {1\over n(n+1)}\oint (-W)^{n+1} =
\oint {1\over W'}\left[{1\over n(n+1)}\left({(-W)^{n+2}\over
-(n+2)}\right)'\right] }

The $N$-point correlation function of one negative and $N-1$ positive
tachyons follows by differentiating this equation. We find
\eqn\nptfn{ \l\l T_{-n} T_{k_2} \cdots T_{k_N}\r\r_0 =
\prod_{j=1}^{N-3}(j-n)\oint {1\over W'}\left({(-W)^{n-N+3}\over
-(n-N+3)}\right)' \del_{t_{k_2}}W\cdots\del_{t_{k_N}}W }
The especially simple form of correlators with a single negative
tachyon will enable us, later on, to provide an interpretation of this
expression as the integral of a top form on the $(N-3)$-dimensional
moduli space of the $N$-punctured sphere.

The other three-point functions in the small phase are the ones with
two or three negative tachyons:
\eqn\twoorthree{
\eqalign{
\l\l T_{-n_1}(t)T_{-n_2}(t)T_{k_3}\r\r_0 &=
\oint {1\over W'}\;
\left({(-W)^{n_1}\over -n_1}\right)'_-\;
\left({(-W)^{n_2}\over -n_2}\right)'_-\;
{\del W\over\del t_{k_3}}\cr
\l\l T_{-n_1}(t)T_{-n_2}(t)T_{-n_3}\r\r_0 &=
\oint {1\over W'}\;
\left({(-W)^{n_1}\over -n_1}\right)'_-\;
\left({(-W)^{n_2}\over -n_2}\right)'_-\;
\left({(-W)^{n_3}\over -n_3}\right)'_- \cr}}
With this information, one can write the superpotential in the big
phase space, where the negative tachyons $T_{-k}$ are also taken to
generate perturbations with parameters $\tbar_k$:
\eqn\bigps{
\eqalign{
W(t,\tbar) = - X^{-1} &+ \sum_{k>0} t_k X^{k-1} +
\sum_{k>0} \tbar_k \left({(-W(t))^k\over -k}\right)'_-\cr
&+ \half \sum_{k_1,k_2 >0}
\tbar_{k_1} \tbar_{k_2} \left( { \left({(-W(t))^{k_1}\over -k_1}\right)'_-
\left({(-W(t))^{k_2}\over -k_2}\right)'_-\over W'(t)} \right)'_-
+ {\cal O}(\tbar^3)\cr}}

Remarkably, it turns out that to this order, the above expression for
$W(t,\tbar)$ is equivalent to
\eqn\equivto{
W(t) = - X^{-1} + \sum_{k>0} t_k X^{k-1} - {1\over\mu^2}
\sum_{k>0} k \l\l T_k \r\r X^{-k-1}}
where now the $\tbar$ dependence is implicit in the last term. To see,
this, differentiate both sides in $\tbar$ and set $\tbar=0$ to get
\eqn\differ{
\eqalign{
{\del\over \del\tbar_n} W \big |_{\tbar=0} &= - \sum_{k>0} k X^{-k-1}
\l\l T_k T_{-n} \r\r \cr
&= -\sum_{k>0} k X^{-k-1} \oint {(-W(Y))^n\over -n} Y^{k-1} dY\cr
&= -\oint \left({(-W(Y))^n\over -n}\right)_- {1\over (X-Y)^2} dY\cr
&= \left({(-W(X))^n\over -n}\right)'_- \cr}}
as expected. Similarly, differentiating twice gives the last term in
Eq.\bigps.

Let us introduce the partition function of the theory in the big phase
space,
\eqn\partition{
Z(t,\tbar) = e^{\mu^2 F(t,\tbar)},}
where $F(t, \tbar)$ is the free energy, and $\mu$ the cosmological
constant. $F(t,\tbar)$ has a genus expansion, $F=
\sum_{g=0}^{\infty}{1\over \mu^{2g}} F_g$, and in this section we
restricted ourselves to the sphere contribution $F_0$ of
order zero in $\mu$.

Since $\l T_k \r = {\partial \over \partial t_k} F$,
Eq.\equivto\ implies that the perturbed superpotential, when
multiplying  $Z(t,\tbar)$ can be substituted by the operator-valued
superpotential ${\What}$, at lowest order in $1/mu^2$:
\eqn\quantized{
\What (t) = -X^{-1} + \sum_{k>0} t_k X^{k-1} -
{1\over\mu^2}\sum_{k>0} k {\del\over \del
t_k} X^{-k-1}.}
In this form, we say that the superpotential is ``quantized''. The
expression for $\What (X)$ is precisely the mode expansion of a
free spin-1 current in a conformal invariant field theory on the
complex plane whose complex coordinate is the Landau-Ginzburg
superfield $X$. (This fact looks very natural if, as suggested in
Ref.\gm, the LG superpotential is identified with the quantized string
field.)

We can succintly rewrite the Eqs. \onept\ and \twoorthree\ for the
correlation functions on the sphere, in the small phase space,
with at most three negative tachyons, in terms of the ``quantized''
superpotential as follows:
\eqn\ward{{1\over \mu^2}{\partial \over \partial\tbar_n}Z =
{1\over n(n+1)}\oint :(-\What)^{n+1}:Z.}
This is the matrix model result in Ref. \dmp\ restricted to genus
zero. We have just seen that the flow equations \flow\ in the small
phase space (which originate from the contact term prescription in the
unperturbed theory) are sufficient to demonstrate Eq.\ward\ up to
third order in $\tbar$ in the free energy $F_0(t,\tbar)$. In Ref.\hop,
it has been shown that one can prescribe an infinite set of
multi-contact terms, such that Eq.\equivto\ is true to all orders in
$\tbar$. This in turn implies that, in genus-0, Eq.\ward\ is true with
$\What$ given by Eq.\quantized.

The contact terms required for this theory (both the
``simple-contact'' appearing in Eq.\flow\ and the multi-contacts
mentioned above) have not so far been actually derived from the
Landau-Ginzburg Lagrangian, but must be taken as an ansatz which gives
a consistent solution. (This is essentially true also of the contact
terms appearing in the polynomial LG theories describing $c<1$ string
backgrounds, which were used implicitly in Ref.\ref\dvv{\DVV}
and explicitly in Ref.\ref\losev{\LOSEV}.) In the present case,
since correct genus-0 amplitudes
incorporating all contact terms can be summarized in the quantization
of the superpotential, it is convenient to assume that quantization of
the superpotential is a basic physical property of this theory. A
more detailed understanding of the compactified moduli spaces relevant
to noncritical strings would be required to provide a rigorous
justification for this ansatz.

In the following sections we will, however, show that this ansatz
satisfies a very stringent consistency check which is derived by
considering the theory at higher genus. In fact Eq.\ward\ is not
consistent as it stands at higher genus, since the operators acting on
the partition function on the R.H.S.,
\eqn\generator{{\cal W}^{(n)}={1\over n(n+1)}\oint :(-\What)^{n+1}:,}
do not satisfy the integrability condition. Their commutators up
to terms of order $1/\mu^4$ are
\eqn\commutator{\left[{\cal W}^{(n)}, {\cal W}^{(m)}\right] =
{1\over 24\mu^4} (n-1)(m-1) (m-n) \oint :(-\What)^{n+m-5}
(-\What^{\prime})^3: +\,\, {\cal O}\left({1\over \mu^6}\right).}
This implies that the higher-genus generalization of
Eq.\ward
\eqn\hgward{{1\over \mu^2}{\partial \over \partial\tbar_n}Z =
\Wop (\mu) Z,}
must involve some deformation $\Wop (\mu)$ of the genus-0 operator
$\Wop$, admitting an expansion of the form
\eqn\deformed{\Wop (\mu) = \Wop + {1\over
\mu^2}\Wopone +  {1\over \mu^4}\Woptwo +
\cdots }
The higher-genus contributions ${1\over \mu^{2g}}\Wopg$
should be such that the deformed operators
$\Wop (\mu)$ commute. However, it is known that, assuming
that the $\Wop (\mu)$ are polynomials in the operator valued
superpotential $\What(X)$ and its derivatives, the
integrability condition determines {\it uniquely}
the higher-genus contributions $\Wopg$, order by order
in ${1\over \mu^2}$.
For example it easy to verify that the genus-1 contribution $\Wopone$
which cancels the commutator in Eq.\commutator\ up to order
${1\over \mu^4}$ is
\eqn\onecontribution{\Wopone = {1-n\over 24}\oint
:(-\What)^{n-2}{\What}^{\prime\prime}:.}
The higher-genus Ward identities \hgward\ have been derived from
matrix-models in Ref. \dmp . Explicit expressions for the operatorial
coefficients $\Wopg$ for any genus $g$ are derived in the Appendix.

In the following sections we will derive the one-point functions of
negative tachyons in the small phase space starting from the
analysis of the LG topological theory on higher-genus world-sheet
surfaces. The higher-genus contributions to the negative tachyon
one-point functions we compute agree precisely with the
ones determined by the Ward identities \hgward\ involving the deformed
operators in Eq.\deformed . We regard this as a higher-genus,
non-trivial verification of the superpotential ``quantization'' ansatz.
Moreover, our results strongly suggest  that the LG topological
theory with the singular superpotential $X^{-1}$ does in fact
define a consistent world-sheet quantum field theory.

\newsec{Picture-changing of Negative Tachyons}

We have already pointed out that expressions for $N$-point functions,
such as Eq.\nptfn\ above, should have an interpretation in a
topological theory as the integral of a top form over the relevant
moduli space. The specially simple form of Eq.\nptfn, where there is a
single negative tachyon, allows us to find the desired interpretation.
This equation is of the form of a matter Landau-Ginzburg
correlator\dvv\ref\vafa{\VAFA} where the positive tachyons
appear in their usual form, but the negative tachyon, in an $N$-point
function, appears as
\eqn\negtachyon{
T_{-n} \sim \prod_{j=1}^{N-3} (j-n)
\left({(-W)^{n-N+3}\over -(n-N+3)}\right)'}
(Recall that the minus-subscript was dropped since it does not
contribute in the presence of only positive tachyons.)
In the absence of perturbations, this field carries an effective
``momentum'' of $-n+N-3$ units. For the special case of three-point
functions, where the moduli space is trivial, the above expression
reduces to the standard one for negative tachyons. Thus we can
interpret Eq.\nptfn\ as saying that, in genus 0, the negative tachyon
appears in many different ``pictures'', with the contribution for a
particular correlator coming entirely from the relevant picture in
which momentum is correctly conserved.

The reason why this picture-changing hypothesis makes sense is
the following: the topological conservation law says that it is
not momentum alone which is conserved, but the difference between
momentum and form dimension, where a gravitational secondary
$\sigma_n$ carries form dimension $n$. Thus in order to view the
negative tachyon as carrying a different momentum in different
pictures, it is necessary to compensate by also assigning it a
different form dimension in each picture, so that the contribution to
the conservation law is the same independent of the picture.

The precise statement of this is that the negative tachyon is really a
sum of gravitational descendants above the fields $T_{-n}$ that we
have been working with above, and reduces to $T_{-n}$ only when the
moduli space is trivial. Thus we write
\eqn\pictures{{\cal T}_{-n} = \sum_{i=0}^n \prod_{j=1}^i (j-n)
\sigma_i(T_{-n+i})}
where the $T_{-n}$ are given in Eq.\negatac\ (without the
minus-subscript, as we have already remarked). Note that the numerical
factor on the RHS of this expression can independently be found\gm\ by
an appropriate continuation of the argument, due to
Losev\losev, that relates matter-sector fields to
gravitational secondaries. This lends further support to the above
identification. Henceforth we interpret this expression as the
physical negative tachyon.

This notion of picture-changing neatly solves the puzzle we raised
above, that the $N$-point function should be a form of dimension $N-3$
integrated over moduli space. The correlation function of one negative
and $N-1$ positive tachyons should now be calculated using ${\cal
T}_{-n}$ above to represent the negative tachyon. Of the various terms
in ${\cal T}_{-n}$, the only one which will contribute is the one in
which on the one hand, the matter sector conserves momentum, while on
the other hand the gravitational sector has the form dimension
appropriate to make it a top form on the moduli space. It is evident
that both requirements are satisfied simultaneously, and by just one
of the terms in Eq.\pictures\ above. Since the positive tachyons are
all gravitational primaries, the relevant sphere correlation function
in the gravitational sector is $\l\sigma_n \sigma_0 \ldots \sigma_0\r$
which is 1 when the moduli space has dimension $n$ and 0
otherwise. Hence the answer reduces to a pure matter-LG calculation
and we manifestly recover Eq.\nptfn.

The above hypothesis should be thought of as the precise version of a
rather general statement made in Refs.\gm\ and \hop\ that the negative
tachyon behaves as a gravitational secondary. In fact, it is the sum
of many gravitational secondaries, of form dimensions varying from 0
to $n$, of which the first (primary) term contributes for sphere
3-point functions, but the other terms contribute on nontrivial moduli
spaces.

As it stands, our proposed picture-changing expression makes sense
only for correlators involving a single negative tachyon. When two or
more negative tachyons are present, multi-point contact terms will be
present and this simple picture will not hold. Fortunately, the fact
that these multi-contact terms are summarised in the quantization of
the superpotential means that it is enough to do everything for
one-point functions of the negative tachyon, in the small phase space
where $\tbar=0$. This will lead to the small-phase-space version of
the $W_{1+\infty}$ constraints of Ref.\dmp, from which one can argue
that the big phase space result follows upon quantization of the
superpotential.

This is not equivalent to suggesting that the equations of the
subsequent section can be extended directly to the big phase space;
they can be so extended only after rewriting them as operator
constraints on the partition function. It would be useful to show
directly that multi-negative-tachyon correlators can be correctly
derived in the LG approach, but this seems rather difficult and
we leave it as an open problem.

\newsec{Higher-genus Correlators}

In higher-genus, the moduli space is always nontrivial. We will use
the expression Eq.\pictures\ and some simple facts about matter LG
systems to obtain the correlation functions. The basic relation we
will need is that if ${\cal O}$ is any operator in a matter LG theory,
then genus-$g$ and genus-0 correlators are related as
follows\vafa,\ref\wittengrav{\WITTENGRAV}:
\eqn\basic{
\l {\cal O} \r_g = \l (W'')^g {\cal O} \r_0 }
where the second derivative of the superpotential, $W''$, can be
thought of as the ``handle operator''.

Our point of view will be that correlators of the LG theory coupled to
gravity can be constructed using the above expression to represent the
handles, and appropriately picture-changed operators to represent the
insertion of the tachyon field. At this stage the computation
factorizes into a matter-like contribution in the form of a
contour-integral (containing the matter part of the picture-changed
tachyon) and the correlation functions of the gravitational operators
$\sigma_n$, which are computed from pure topological
gravity\ref\wittenint{\WITTENINT} (equivalent to intersection theory
on moduli space).

For example, in genus 1 with one puncture, the moduli space is
1-dimensional, and we have
\eqn\genusone{
\eqalign{
\l\l {\cal T}_{-n} \r\r_{g=1}(\tbar=0)
&= \l\sigma_1 \r_{g=1} \oint {W''\over W'}
(1-n) \left({(-W)^{n-1}\over -(n-1)}\right)'\cr
&= {1-n \over 4!} \oint W''(-W)^{n-2} \cr }}
Here we have used the result from Ref.\wittenint\
that $\l \sigma_1 \r_{g=1} = {1\over 4!}$. The above
expression is precisely the torus one-point function of a negative
tachyon as obtained from matrix models, at $\tbar=0$!

Before going on to look at general genus, it is useful to write down
the known answers from matrix models for ease of comparison. The
one-point function of a negative tachyon at $\tbar=0$ is given in
Ref.\dmp\ as a function of the cosmological constant $\mu$. The
genus-$g$ contribution is obtained by expanding in powers of $1/\mu$
and keeping the term of order $\mu^{-2g}$ relative to the tree level
term. This is worked out in the Appendix. The result is a
combinatorial formula involving a sum over partitions of $g$, of the
form $1^{\alpha_1} 2^{\alpha_2} \cdots g^{\alpha_g}$ with
$\sum_{l=1}^g l\alpha_l =g$. Let $p=\sum_{l=1}^g \alpha_l$ be the
total number of elements in the partition, then
\eqn\comb{
\eqalign{
\l\l {\cal T}_{-n} \r\r_g(\tbar=0) &=
\sum_{{\alpha_1,\ldots,\alpha_g
\atop \sum \alpha_l =p,\,\sum l\alpha_l = g}}
\l\l {\cal T}_{-n} \r\r_g^{(\alpha_1,\ldots,\alpha_g)}\cr
\l\l {\cal T}_{-n} \r\r_g^{(\alpha_1,\ldots,\alpha_g)} &=
{1\over 2^{2g}}{\prod_{j=1}^{2g-2+p} (j-n)\over \prod_{l=1}^g
\alpha_l!\left( (2l + 1)!\right)^{\alpha_l}}\cr
& \qquad
\oint {\prod_{l=1}^g (\del^{2l}W)^{\alpha_l} \over W'}
\left({(-W)^{n-2g+2-p}\over -(n-2g+2-p)}\right)' \cr}}
(Here, $\del$ represents $\del/\del X$.)

In particular, the contribution from the specific partition with
$\alpha_1=g, \alpha_2,\ldots =0$, for which $p=g$, is
\eqn\bulk{\l\l {\cal T}_{-n} \r\r_g^{(g,0,\ldots,0)} =
{1\over 2^{2g}} {\prod_{j=1}^{3g-2} (j-n) \over g! (3!)^g} \oint
{(W'')^g\over W'} \left({(-W)^{n-3g+2} \over -(n-3g+2)}\right)'}
but in general there are several other terms, except in genus 1 where
the above term gives the whole answer. {}From this we see in particular
that the genus 1 expression Eq.\genusone\ calculated from LG theory
agrees with the matrix model result.

Recalling Eq.\pictures, one sees that in every term of Eq.\comb\
above, a picture-changed tachyon appears, but each time in a different
picture. Indeed, for a partition into $p$ elements, the tachyon
appears in the $2g-2+p$ picture, so it must be thought of as a
$2g-2+p$ form on moduli space. For the special partition which
contributes to Eq.\bulk, we have $p=g$ and hence the tachyon appears
here in the $3g-2$ picture, which corresponds to the dimension of the
moduli space $\mg1$. We conclude that this is the unique term in the
matrix model amplitude which can be thought of as an integral of a top
form over the moduli space of genus $g$ with one puncture. The other
terms must correspond to boundaries of $\cmg1$.

Returning now to the Landau-Ginzburg theory, we generalize to
arbitrary genus the calculation leading to Eq.\genusone, to find
\eqn\genusg{
\l\l {\cal T}_{-n} \r\r_g(\tbar=0)
= \l\sigma_{3g-2} \r_g \oint {(W'')^g\over W'}
\prod_{j=1}^{3g-2} (j-n) \left({(-W)^{n-3g+2}\over -(n-3g+2)}\right)'}
{}From pure topological gravity it is known\wittenint\ that
\eqn\topgrav{
\l \sigma_{3g-2} \r_g = {1\over g! (4!)^g}}
Inserting this in Eq.\genusg, one finds that this LG result is
precisely equal to Eq.\bulk\ in every genus, including all factors!

Thus the LG theory reproduces the ``bulk'' term in every genus. To get
the other terms, one has to analyse the various boundaries of $\cmg1$
and see whether the extra terms in Eq.\comb\ can be interpreted as
arising from boundary contributions, or in other words as contact
terms between handles. We turn to this in the following section.

\newsec{Boundary of Moduli Space and Handle Contact Terms}

Consider the case of genus-2, which is the first one for which the
bulk contribution to the tachyon amplitude is not the whole answer.
{}From Eq.\comb\ it is easy to work out that the genus-2 amplitude is
\eqn\genustwo{
\eqalign{
\l\l {\cal T}_{-n}\r\r_{g=2} = &{1\over (4!)^2 2}
(1-n)(2-n)(3-n)(4-n)\oint {(W'')^2\over W'}
\left({(-W)^{n-4}\over -(n-4)}\right)'\cr
+ &{1\over 4^2 5!} (1-n)(2-n)(3-n)\oint {W''''\over W'}
\left({(-W)^{n-3}\over -(n-3)}\right)'\cr }}
The second term above is our first example of a boundary term. Clearly
it comes from a boundary, of complex dimension 3, of the dimension-4
moduli space $\overline{{\cal M}}_{2,1}$.

This leads us to postulate contact terms between the handle operators
$W''$ of the Landau-Ginzburg theory, analogous to the contact terms
between tachyon operators\losev\ that were crucial in Refs.\gm,\hop\ to
define the theory consistently. We have mentioned at the end of
section 2 that the handle contact terms are essential for the ${\cal
W}^{(n)}$ generators to commute, hence they are indeed required by
consistency.

To start with, we propose that the contact term between a pair of
handles is
\eqn\contact{
C(W'',W'') = \alpha~W'''' }
where $\alpha$ is a numerical constant. We call this a `simple
contact', since we will see that there are also multiple contacts.
Geometrically we imagine that a contact between a pair of handles
arises when two handles coincide, so that a nontrivial homology cycle
is pinched. The surface acquires two extra punctures and reduces its
genus by one, hence the complex dimension of the moduli space goes
down by one as desired.

In genus 2, the contact term above will therefore give a contribution
\eqn\contacttwo{
\l\l {\cal T}_{-n}\r\r_{g=2}^{contact} = \l \sigma_3 \sigma_0 \sigma_0
\r_{g=1}
\oint {C(W'',W'')\over W'} (1-n)(2-n)(3-n) \left({(-W)^{n-3}\over
-(n-3)}\right)' }
where the tachyon has appeared in the ``3''-picture. The two
$\sigma_0$ operators are associated to the two extra punctures on the
collapsed surface, which has genus 1. Using the recursion relations of
pure topological gravity (in particular, the string equation) we find
\eqn\sigmathreept{
\l\sigma_3 \sigma_0 \sigma_0\r_{g=1} = \l \sigma_1\r_{g=1} = {1\over 4!}}
Inserting this and the contact term in Eq.\contacttwo, we get
\eqn\contactthree{
\l\l {\cal T}_{-n}\r\r_{g=2}^{contact} =
{\alpha\over 4!} (1-n)(2-n)(3-n)\oint {W''''\over W'}
\left({(-W)^{n-3}\over -(n-3)}\right)' }
Comparing with Eq.\genustwo\ determines the arbitrary constant
appearing in the contact term to be
\eqn\coeff{
\alpha= {1\over 5.2^4}}

As a check, we now reproduce the boundary terms in arbitrary genus
associated to the contact of a single pair of handles. This is
equivalent to pinching of a single nontrivial homology cycle,
connecting any two handles. There are $g-1$ such cycles, and the
associated partition of $g$ in Eq.\comb\ is clearly
$\alpha_1=g-2,\alpha_2=1,\alpha_3=\cdots=0$. Thus we find the
contribution
\eqn\simplecontact{
\langle \sigma_{3g-3} \sigma_0 \sigma_0 \rangle_{g-1}
\prod_{j=1}^{3g-3}(j-n)\oint
(g-1){C(W'', W'') (W'')^{g-2}\over W'}
\left({(-W)^{n-3g+3}\over -(n-3g+3)}\right)' }
Using the string equation, which gives
\eqn\stringeqn{
\langle \sigma_{3g-3} \sigma_0 \sigma_0 \rangle_{g-1} = \langle
\sigma_{3g-5} \rangle_{g-1} = {1\over (4!)^{g-1} (g-1)!}}
the contribution becomes
\eqn\contribution{
{1\over 2^{2g}}
{\prod_{j=1}^{3g-3}(j-n)\over 5! (3!)^{g-2} (g-2)!}
\oint {W''''(W'')^{g-2}\over W'}
\left({(-W)^{n-3g+3}\over -(n-3g+3)}\right)' }
which is seen to agree with
$\l\l {\cal T}_{-n}\r\r_g^{(g-2,1,0,\ldots,0)}$ of Eq.\comb\ above.

Thus we have found that a simple handle-contact term reproduces a
class of contributions to the tachyon amplitude in every genus,
corresponding to the pinching of a single non-trivial cycle. Suppose
now we allow the pinching of two such nontrivial cycles, then there
are two ways that this can happen: either the two pinched cycles link
two completely disjoint pairs of handles, or they link two adjacent
pairs of handles (so one handle is common). In the first case, the
simple contact term suffices to describe this effect, and leads to
answers containing two factors of $W''''$, with some fixed
coefficients which can be computed. In the second case, we have a
fusing of three handles, and will have to postulate a new contact term
to describe this.

Let us look at the first case in more detail. In this case, we are
concerned with a boundary of complex codimension 2. To do the
counting, we must first assume that we have pinched one of the
relevant cycles, and within this boundary (of codimension 1) we have
to count the number of ways in which a second cycle can be pinched.
Requiring that the second cycle be disjoint from the first, it can
link any two out of $g-2$ possible handles, so the combinatorial
factor is ${(g-2)(g-3)\over 2}$. (In the second case, to which we
return later, the region is again a boundary of codimension 2, and a
similar counting gives a factor $g-2$.) With this factor, it is
straightforward to show that the term
$\l\l {\cal T}_{-n}\r\r_g^{(g-4,2,0,\ldots,0)}$ of Eq.\comb\ above
is reproduced by using a pair of simple contacts. This can be further
generalized to all terms corresponding to partitions of the type
$(g-2m,m,0,\ldots,0)$, describing the pinching of $m$ disjoint cycles.

To reproduce partitions for which $\alpha_l\ne 0$ for $l\ge 3$, we
need multi-contact terms which represent the fusing of 3 or more
handles. For example, the multi-contact betwen three handles produces
a term with 6 derivatives of the superpotential. Following the same
procedure as for the simple contact, one finds
\eqn\threecontact{
C(W'',W'',W'') = {4!\over 7!\, 2^6} W''''''}
The interpretation of this term is that it comes from the boundary of
moduli space where two adjacent non-trivial cycles are pinched. In
such terms, the tachyon will appear in the picture relevant to the
dimension of this boundary, which is $3g-4$, which is of complex
codimension 2 just as for the pinching of non-adjacent cycles. This
agrees with the analysis of the appropriate terms in Eq.\comb\
containing the sixth derivative of the superpotential.

More generally, the multi-contact between $n$ handles is
\eqn\multicontact{
C(W'',W'',\ldots,W'')~(n~\hbox{\rm times}) =
{4!\over (2n+1)!\, 2^{2n}} \del^{2n} W }
It is easy to see that this correctly reproduces, for example, the
term in genus-$g$ where all the $g$ handles collide, degenerating to a
genus-1 surface. In Eq.\comb, pick the term in genus-$g$ with
$\alpha_g=1$ and all the other $\alpha_i=0$. The $4!$ in
Eq.\multicontact\ above is cancelled by the contribution from the
gravitational sector:
\eqn\gravit{
\l \sigma_{2g-1} (\sigma_0)^{2g-2} \r_{g=1} = {1\over 4!} }
and it is clear that the other factors in the mtulicontact term are
precisely the desired ones.  Note that for $n=1$ the numerical
coefficient on the RHS of Eq.\multicontact\ reduces to unity, as one
would expect since in this case there is no contact.

To conclude, the picture-changing operation that we have described,
along with the formula Eq.\basic\ for matter Landau-Ginzburg
correlators, and the handle-contacts discussed above, reproduce
completely the formula Eq.\comb\ for the negative-tachyon 1-point
functions, at $\tbar=0$, in every genus.

This means that with the assumptions and computations described above,
the formula of Ref.\dmp\ restricted to $\tbar=0$ (Eq.\comb\ above,
which is shown in the Appendix to be equivalent to Eq.(8.1)) has been
obtained in our topological LG framework.  With the quantized
superpotential, Eq.(8.1) should be viewed as an operator acting on the
partition function of Eq.\partition\ above, in which case it gives all
correlators for arbitrary configurations of positive and negative
tachyons. This completes the chain of arguments to the effect that the
$c=1$ matrix model results follow from the topological Landau-Ginzburg
formulation, modulo the remarks at the end of Section 3.

\newsec{Partition Function}

Having obtained the tachyon correlators in every genus from
Landau-Ginzburg considerations, we may ask whether this also enables
us to find the partition function in each genus. We now show that this
is indeed the case.

Although Eq.\comb\ is strictly true for negative tachyons, namely
$n\ge 1$, one can check if its limit as $n\to 0$ makes sense. In the
original form of Eq.(8.1) of the Appendix (which came from matrix
models) this limit appears rather singular. However, in the equivalent
form of Eq.\comb\ which was eventually derived from Landau-Ginzburg
theory, it is quite easy to take the limit. Restricting to the
unperturbed superpotential $W=-1/X$, one finds
\eqn\limitofcomb{
\l\l {\cal T}_0 \r\r_g = {1\over 2^{2g}}
\sum_{{\alpha_1,\ldots,\alpha_g
\atop \sum \alpha_l =p,\,\sum l\alpha_l = g}}
(-1)^p {(2g-2+p)! \over \prod_{l=1}^g
\alpha_l!(2l + 1)^{\alpha_l}} }

Since $T_0$ is the cosmological operator\gm, this is equal to
$\del/\del\mu(Z(\mu))$ expanded in powers of $1/\mu$, where
\eqn\partfnexp{
Z(\mu) = \sum_{g=0}^\infty Z_g \mu^{2-2g}}
Thus the RHS of Eq.\limitofcomb, which we denote by $A_g$, equals
$(2-2g)Z_g$. This gives us an explicit way to compute $Z_g$.

We now show that $A_g$ is equal to $- B_{2g}/2g$ where $B_{2g}$ are
the Bernoulli numbers. Let us write
\eqn\zedexp{
A_g =  \sum_{p=1}^g A_{g,p} (2g-2+p)!,}
which defines the $A_{g,p}$.
The corresponding generating function
\eqn\generating{
A(z,\lambda) \equiv \sum_{g=0}^\infty \sum_{p=1}^g A_{g,p} z^p
\lambda^{2g}}
becomes
\eqn\gencalc{
\eqalign{
A(z,\lambda) &= \sum_{g,p} (-z)^p \left( {\lambda\over2}\right)^{2g}
\sum_{{\alpha_1,\ldots,\alpha_g
\atop \sum \alpha_l =p,\,\sum l\alpha_l = g}} {1\over \prod_{l=1}^g
\alpha_l!(2l + 1)^{\alpha_l}}\cr
&= \prod_{l=1}^\infty \sum_{\alpha_l=0}^\infty {1\over\alpha_l!}
\left( -{z\over 2l+1}\left({\lambda\over2}\right)^l\right)^{\alpha_l}\cr
&= \exp\left(-z \sum_{l=1}^\infty {1\over 2l+1} \left( {\lambda\over
2}\right)^l\right)\cr
&= e^z \left( {1- \lambda/2 \over 1+\lambda/2}
\right)^{z/\lambda}.\cr}}
Now the idea is to use the Borel resummation trick to obtain the
generating function $A(\lambda)$ for the $A_g$,
starting from the known $A(z;\lambda)$:
\eqn\resum{
\eqalign{
A(\lambda)\equiv & \sum_{g=0}^\infty \lambda^{2g} A_g \cr
=& \int_0^{\infty} dt e^{-t} \sum_{g=0}^\infty \sum_{p=1}^g t^{2g-2+p}
\lambda^{2g} A_{g,p} \cr
=&\int_0^{\infty} {dt \over t^2} e^{-t} A(t;\lambda t)\cr
=& \int_0^{\infty} {dt \over t^2} \left( {1- \lambda t/2 \over
1+ \lambda t/2}\right)^{1/\lambda},\cr}}
Performing the substitution
\eqn\subst{
e^{-x} = {1- \lambda t/2 \over 1+ \lambda t/2},}
one obtains :
\eqn\alamb{
A(\lambda) = {\lambda \over 4} \int dx e^{-x/ \lambda} {1\over
{\sinh}^2 (x/2)}.}
Making the identification $\lambda = 1/i\mu$, one verifies that
the expansion in inverse powers of $\mu$ of this expression is
precisely $-B_{2g}/2g$ (see Eqs. (5.36),(5.37) of Ref.\dmp).

Thus we have proved that $A_g = -B_{2g}/2g$, so that
\eqn\finalpart{
Z_g = {B_{2g}\over 2g(2g-2)} }
which is precisely the partition function originally obtained from
matrix models compactified at the self-dual
radius\ref\grosskleb{\GROSSKLEB}.

\newsec{Discussion and Conclusions}

We have considered the topological Landau-Ginsburg version of 2d
string theory at higher genus. The main result of our analysis is that
the higher-genus tachyonic correlation functions known from matrix
models admit a topological decomposition as a sum of contributions
from the interior and from the boundary of moduli space. We also
showed that the bulk contribution is easily computed, at all genus, by
combining results from topological matter field theory and topological
gravity. The contributions from the boundaries of moduli space have
been shown to be interpretable as contact terms between collapsing
handles.  We also extended our topological interpretation of 2d string
correlators to the partition function, which at each genus is given by
the virtual Euler characteristic of moduli space.

Our work brings the $c=1$ string into the same topological framework
as the $c <1$ strings. Similarly to the $c<1$ models, the contribution
to the correlation functions coming from the interior of the moduli
space is easily determined from topological field theory.  As for the
contributions from the boundary of the moduli space, they are
determined from consistency rather than from strictly field
theoretical methods. It would be interesting if these latter
contributions could be explicitly re-derived by extending the analysis
of Ref.\vafa\ to degenerated Riemann surfaces.

Two features of the topological resolution of the 2d string that we
propose appear to be novel. First, it turns out that some of the
physical operators of the theory -- the negative tachyons -- should be
picture-changed, not only near the boundary of moduli spaces as in
topological gravity\dvv, but in the bulk part of the correlators as
well. Second, the kind of degenerations of Riemann surfaces
which give rise to the relevant boundary terms -- multi-contacts between
collapsing handles -- seem to correspond to a compactification of the
moduli space which is not the same as the one appearing in topological
gravity.  It would be interesting to understand both these features
from a genuinely field theoretical point of view.

\newsec{Appendix: Genus Expansion of $W_\infty$ Constraints}
In this appendix, we will show that if the generating function of
Ref.\dmp\ at $\tbar=0$:
\eqn\dmponeptfn{
\l\l T_{-n}\r\r(\tbar=0) =
\oint{1\over n(n+1)}e^{-i\mu \phi (X)} \left({\del
\over i\mu}\right)^{n+1} e^{i\mu \phi (X)}}
is expanded in inverse powers of $\mu$ as follows:
\eqn\gexpansion{
\l\l T_{-n}\r\r = \sum_{g=0}^\infty (i\mu)^{-2g}\l\l
T_{-n}\r\r_g}
then $\l\l T_{-n}\r\r_g(\tbar=0)$ is given by Eq.\comb\ above, where
$W(X)=-\del\phi(X)$ has the mode expansion in Eq.\potential.

In order to prove Eq.\comb\ we will establish the following,
more general result. The operators $\Wop (\mu)$ appearing in
the Ward identities \hgward\ derived from matrix-models in
Ref.\dmp\ are
\eqn\defoperator{\Wop (\mu) = {1\over n(n+1)}:e^{-i\mu \phihat (X)}
\left({\del \over i\mu}\right)^{n+1} e^{i\mu \phihat (X)}:}
where $\What (X) = - \del \phihat (X)$. We will show that
the operators $\Wopg$ defined by the expansion \deformed\
are given by a formula analogous to Eq.\comb :
\eqn\opcomb{\eqalign{
\Wopg =&{1\over 2^{2g}}{\prod_{j=1}^{2g-2+p} (j-n)\over \prod_{l=1}^g
\alpha_l!\left( (2l + 1)!\right)^{\alpha_l}}\cr
&\qquad
\oint :{\prod_{l=1}^g (\del^{2l}\What )^{\alpha_l} \over {\What}'}
\left({(-\What )^{n-2g+2-p}\over -(n-2g+2-p)}\right)': \cr}}

When considering the Ward identities \hgward\ at  $\tbar=0$,
one can simply substitute the quantized superpotential $\What (X)$ with
the ``classical'' superpotential $W(X)$ of Eq.\potential\
in the operatorial expression for $\Wopg$. This is because the derivative
terms in Eq.\quantized\ pull down positive-momentum tachyon
correlators from the partition function, but these all vanish at
$\tbar=0$. Thus, Eq.\opcomb\ implies, in particular, the validity
of Eq.\comb .

Introduce the generating operator-valued field $\Wcal(z;\mu)$ defined by
\eqn\tzmusum{
\Wcal (z;\mu) \equiv \sum_{n=0}^\infty {z^{n+1}\over (n-1)!}\Wop (\mu)}
Subsituting Eq.\defoperator\ in the definition above, and using the
Taylor expansion formula, we can write it as
\eqn\tzmuint{
\Wcal (z;\mu) = \oint :e^{i\mu \phihat(X + z/i\mu) -i\mu\phihat(X)}: }

Since $\Wcal (z;\mu)$ is an even function of $\mu$, it is convenient to
make this manifest by shifting $X \to X - z/ 2i\mu$:
\eqn\defines{
\Wcal (z;\mu) =
\oint :e^{i\mu (\phihat(X + z/ 2i\mu) -\phihat(X - z/2i\mu))}:~
\equiv \oint :\exp\left(i\mu S(X; z/ i\mu)\right):~,}
The ``action'' $S(X; z/ i\mu)$ can be expanded as
\eqn\action{
S(x; z/ i\mu)= \sum_{l=0}^\infty {2\over(2l+1)!} \left(
{z\over 2i\mu}\right)^{2l+1} \del^{2l+1}\phihat(X). }
which gives the generating field in the form:
\eqn\tzmufinal{
\Wcal (z;\mu) = \oint :e^{z \del\phihat(X)} \exp\left[\sum_{l=1}^\infty {z\over
(2l+1)!2^{2l}} \left({z \over i\mu}\right)^{2l}
\del^{2l+1}\phihat(X)\right]: }

We now show that Eq.\opcomb\ leads to the same generating field, thereby
proving its equivalence to Eq.\defoperator . Inserting this in
Eq.\gexpansion\ and Eq.\tzmusum, and performing first the sum over
$n$, we get
\eqn\intermed{
\eqalign{
\sum_{n=0}^\infty{z^{n+1}\over (n-1)!}
\prod_{j=1}^{2g-2+p}(j-n) :(-\What )^{n-2g+1-p}:
&= \left({\del\over\del \What}\right)^{2g+p} \sum_{n=0}^\infty{z^{n+1}\over
(n+1)!}:(-\What )^{n+1}:\cr
&= (-z)^{2g+p} :e^{-z\What }:\cr} }
Using the above in Eq.\tzmusum, we find
\eqn\tzmualt{
\Wcal (z;\mu) = \oint :e^{-z\What }\sum_{g=0}^\infty
\left({z\over 2i\mu}\right)^{2g}
\sum_{{\alpha_1,\ldots,\alpha_g
\atop \sum l\alpha_l = g}}
\prod_{l=1}^g { \left(- z\, \del^{2l+1}\What \right)^{\alpha_l}\over
\alpha_l!\left((2l + 1)!\right)^{\alpha_l}}:  }

One can now use the simple combinatorial identity
\eqn\identity{
\exp\left(\sum_{l=1}^\infty a_l\lambda^l\right) = \sum_{g=0}^\infty
\lambda^g\sum_{{\alpha_1,\ldots,\alpha_g
\atop \sum l\alpha_l = g} }
\prod_{l=1}^g {(a_l)^{\alpha_l}\over \alpha_l!} }
in Eq.\tzmualt\ to show that it is equivalent to the expression
\tzmufinal\ above, with the identification $\What (X)=-\del\phihat (X)$. Thus
we have shown that Eq.\comb\ represents the genus-expansion of the
matrix model results.

\noindent{\bf Acknowledgements}

We are grateful to D. Jatkar, A. Losev, G. Moore, S. Panda, E.
Verlinde and H. Verlinde for helpful discussions. D.G. acknowledges the
International Centre for Theoretical Physics, Trieste; C.I. and S.M.
acknowledge the Laboratoire de Physique Theorique, Ecole Normale
Superieure, Paris, and S.M. is grateful to the Mehta Research
Institute, Allahabad, and Theory Division, CERN, Geneva, for
their kind hospitality at various stages of this work.

\listrefs
\end